\documentclass[a4paper,fleqn,usenatbib]{mnras}

\usepackage{newtxtext,newtxmath}

\usepackage[T1]{fontenc}
\usepackage{ae,aecompl}

\usepackage{graphicx}	
\usepackage{amsmath}	
\usepackage{amssymb}	

\title[The direct detection of SDSS1411+2009B]{The direct detection of the irradiated brown dwarf in the white dwarf - brown dwarf binary SDSS\,J141126.20+200911.1}
\author[S. L. Casewell et al.]{S. L. Casewell$^{1}$ \thanks{E-mail: slc25@le.ac.uk}, S. P. Littlefair$^{2}$, S. G. Parsons$^{2}$, T.R. Marsh$^{3}$, J. J. Fortney$^{4}$, 
\newauthor and M. S. Marley$^{5}$ \\
$^{1}$Department of Physics and Astronomy, University of Leicester, University Road, Leicester LE1 7RH, UK \\
$^{2}$Department of Physics and Astronomy, University of Sheffield, Sheffield, S3 7RH, UK\\
$^{3}$Department of Physics, University of Warwick, Gibbet Hill Road, Coventry CV4 7AL, UK\\
$^{4}$Department of Astronomy and Astrophysics, University of California, Santa Cruz, CA 95064, USA\\
$^{5}$NASA Ames Research Center, MS-245-3, Moffett Field, CA 94035, USA }
\date{Accepted XXX. Received YYY; in original form ZZZ}

\pubyear{2017}

\begin{document}
\label{firstpage}
\pagerange{\pageref{firstpage}--\pageref{lastpage}}
\maketitle

\begin{abstract}
We have observed the eclipsing, post-common envelope white dwarf-brown dwarf binary, SDSS141126.20+200911.1, in the near-IR with the HAWK-I imager, and present here the first direct detection of the dark side of an irradiated brown dwarf in the $H$ band, and a tentative detection in the $K_s$ band. Our analysis of the lightcurves and indicates that the brown dwarf is likely to have an effective temperature of 1300 K, which is not consistent with the effective temperature of 800 K suggested by its mass and radius. As the brown dwarf is already absorbing almost all the white dwarf emission in the $K_s$ band we suggest that this inconsistency may be due to the
UV-irradiation from the white dwarf inducing an artificial brightening in the $K_s$ band, similar to that seen for the similar system WD0137-349B, suggesting this brightening may be characteristic of these UV-irradiated binaries.

\end{abstract}

\begin{keywords}
brown dwarfs, binaries:eclipsing, white dwarfs,
\end{keywords}


\section{Introduction}
Despite recent results reporting the discovery of brown dwarf companions to main sequence stars (e.g. \citet{anderson11,siverd12,bayliss2017,triaud18}), there are still only thirteen known to date, and they are very rare compared to planetary or stellar companions to main sequence stars \citep{grether06, metchev04}.  As a result, there are very few systems known to have evolved from these binaries with  \citet{steele11} predicting only 0.5\% of white dwarfs having brown dwarf companions. 

To date only nine
post-common envelope systems have been confirmed: GD1400 (WD+L6, P=9.98\,hrs; \citealt{farihi04, dobbie05, burleigh11}), WD0137-349 (WD+L6-L8, P=116\,min; \citealt{maxted06, burleigh06}), WD0837+185 (WD+T8, P=4.2\,hrs; \citealt{casewell12}), NLTT5306 (WD+L4-L7, P=101.88\,min; \citealt{steele13}), SDSS J155720.77+091624.6 (WD+L3-L5, P=2.27\,hrs; \citealt{farihi17}), SDSS J1205-0242 (WD+L0, P=71.2\,min; \citealt{parsons17, rappaport17}), SDSS J1231+0041 (WD+M/L, P=72.5\,min; \citealt{parsons17}), EPIC212235321 (WD+L5, P=68 min; \citealt{casewell18}) and SDSS J141126.20+200911.1, hereafter SDSS1411+2009 (WD+T5, P=121.73\,min; \citealt{beuermann13, littlefair14}). All of these systems have survived a phase of common-envelope evolution, resulting in the close binary system. They are all detached, and likely tidally-locked, resulting in a brown dwarf that is irradiated on one hemisphere, similar to the situation in most hot Jupiter exoplanets. Eventually, these white dwarf-brown dwarf binaries will become cataclysmic variables, such as SDSS1433+1011 in which the substellar donor was recently detected \citep{hernandez16}.

Irradiated brown dwarfs are expected to have very similar atmospheres to irradiated exoplanets, and have been described as the "fourth corner" of the parameter space containing irradiated exoplanets, solar system planets and isolated brown dwarfs \citep{showman}. For instance, Kelt-9b \citep{gaudi17} is a 2.88 M$_{\rm Jup}$ planet orbiting a $\sim$10 000 K star. This planet is expected to receive $\sim$700 times more UV irradiation than a planet orbiting the next hottest exoplanet host star (WASP-33). However, the primary star in the Kelt-9 system is still $\sim$3000 K cooler than SDSS1411+1011A and $\sim$6500 K cooler than WD0137-349A. The brown dwarf companion in this latter system  has been shown to have an atmosphere that is significantly affected by UV irradiation \citep{longstaff17, casewell15}. In fact, Kelt-9b has been shown to have a day-nightside temperature difference of $\sim$500 K, the same as the irradiated brown dwarf WD0137-349B, indicating poor heat redistribution is present in both systems, despite their differences in internal temperature.  Studying irradiated brown dwarfs can therefore provide a useful proxy for exoplanet systems, especially to explore the effects of UV irradiation and any resultant photochemistry, as in general hot Jupiter host stars replicating the same conditions must be very large, making them challenging systems to observe. 
One of the most recently discovered of the post-common envelope systems, and the first eclipsing system to be discovered, 
SDSS1411+2009, was discovered as part of the Catalina Sky Survey by \citet{drake10}. The substellar nature of the companion to the white dwarf was confirmed by \citet{beuermann13}. While its period is very similar to that of the well-studied WD0137-349, the white dwarf is cooler with T$_{\rm eff}$=13000$\pm$300 K and log g=7.86$\pm$0.07, giving a mass of 0.53$\pm$0.03 M$_{\odot}$ \citep{littlefair14}. The brown dwarf mass is calculated to be 50$\pm$2 M$_{\rm Jup}$, and has an estimated spectral type of T5, derived from the secondary's mass. The $z'$ band eclipse and $K_s$ excess presented in \citet{littlefair14} were used to estimate the dayside spectral type to be between L7 and T1, suggesting significant irradiation.

\section{Observations and data reduction}

We observed SDSSJ1411+2009 with the infrared imager HAWK-I \citep{kissler08} on the VLT as part of programme 94.C-0032. The data were obtained on the nights of the 2015-04-04, 2015-04-05, and the 2015-03-13 for $J$, $H$ and $K_s$ respectively. The seeing was 1" in the $J$, and $H$ bands and between 1.5" and 2.5" in the $K_s$ band. We used the fast photometry mode, allowing us to window the detector and reduce the deadtime between frames to a few microseconds, and used exposure times of 5 s in each of the $J$, $H$, and $K_s$ bands.  We observed using chip 4, and orientated the 128 pixel window to 120 degrees to also observe a standard star, 2MASS14112391+2008132 which was used to calibrate the photometry. The data were dark-subtracted, flat fielded and extracted using aperture photometry within the ULTRACAM pipeline \citep{dhillon07}. 

\section{Results}

We used \textsc{lroche}, part of the \textsc{lcurve} software to model the lightcurves (see \citealt{copperwheat10} for a description). We sample the posterior probability distributions for model parameters using affine-invariant Markov-chain Monte-Carlo (MCMC) \citep{foreman-mackey2013}. We used the system parameters given in \citet{beuermann13} and \citet{littlefair14} to set priors on the mass ratio, orbital period, angle of inclination, white dwarf temperature and stellar radii. The covariance matrix from \citet{littlefair14} was used to create multi-variate normal priors for the stellar radii and the inclination. Independent Gaussian priors were used for all other parameters. Since the lightcurves show evidence for red noise, presumably arising from instrumental systematics, we do not use the chi-squared statistic to estimate the likelihood. Instead we model the residuals from the \textsc{lroche} model using a Gaussian process with a Mat{\'e}rn-3/2 kernel and use the likelihood of the residuals \citep[see][for an example of this approach]{mcallister17}. Multiple, independent MCMC chains are run from different starting points, and we use the Gelman-Rubin diagnostic, applied to the independent runs, to test for convergence. We also tested that the results were insensitive to the kernel function adopted for the Gaussian process.

We adopt the limb darkening coefficients in \citet{gianninas13} for a 13000 K, log g =8.00 white dwarf for the $y$ band, as there are none available for the near-IR, although as this is within the Raleigh-Jeans tail of the white dwarf spectrum, these coefficients are not expected to deviate much from these values. Additionally, given the S/N of our data, any deviation will have a negligible effect on our fit.

\begin{figure*}
\begin{center}
\scalebox{0.75}{\includegraphics[]{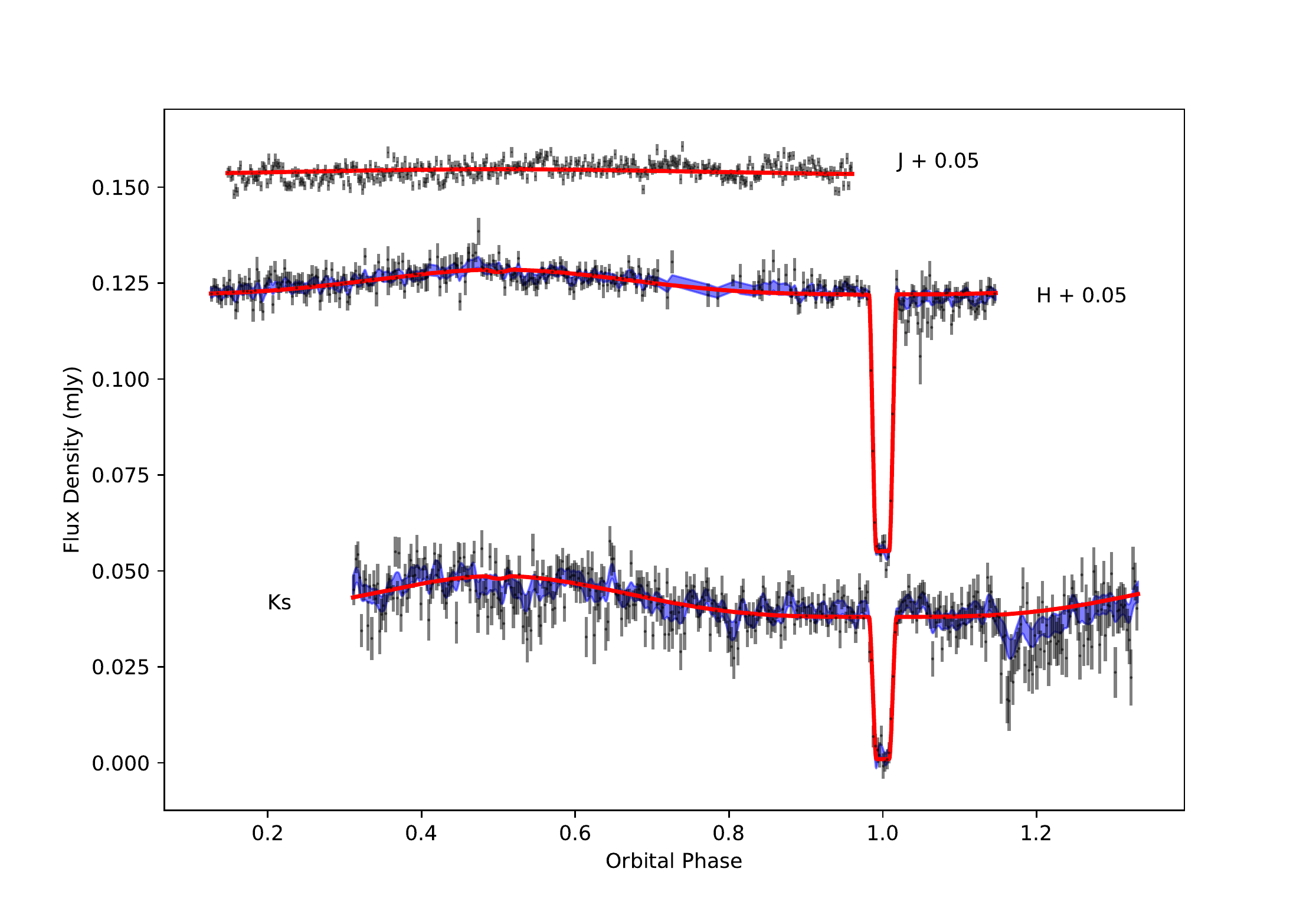}}
\caption{\label{lc}$JHK_s$ band lightcurves phased on the ephemeris in \citet{beuermann13}. The lightcurves have been offset for display purposes by 0.05 mJy in the $H$ and $J$ bands. 
The data have been plotted in 400 error-weighted flux bins for display purposes and the models are plotted with the red solid line. The shaded-blue region represents the 1$\sigma$ prediction of the binary model, plus the Gaussian process model of the systematics.}
\end{center}
\end{figure*}

The {\sc lroche} model is used to measure the level of the reflection effect caused as the heated side of the brown dwarf moves into view. The brightness temperature of an element on the companion is modeled as:
\begin{equation*}
T_{c,j}^4 =  \left[ T_c \left( \frac{g_j}{g_{{\rm pole}}}\right)^{\beta} \right]^4 +  \alpha G_j T_{\rm wd}^4,
\end{equation*}
where $\alpha$ is the fraction of the incident flux which is absorbed (i.e $\alpha = 1 - A$), where  $A$ is the albedo. $g_j$ is the surface gravity of the element, $g_{\rm pole}$ is the surface gravity at the pole, $\beta$ is the gravity darkening exponent, for which we adopted a value of 0.45.  $G_j$ is a geometric factor which accounts for the fraction of the WD flux absorbed by the companion, taking the full Roche geometry into account. $T_c$ and $T_{\rm wd}$ are the black-body brightness temperatures of the companion and white dwarf respectively. Because our observations are within the Raleigh-Jeans tail of the white dwarf spectrum, the surface brightness of a white dwarf differs from that of the same-temperature black-body by less than 5\%.  The lightcurve of an irradiated binary in a single band constrains the {\em ratio} of brightness temperatures of the two components. Therefore, since a black-body is a reasonable description for the white dwarf, we can say that using $T_c$ in the Planck function gives an accurate prediction of the surface flux of the brown dwarf; these surface fluxes can be compared directly with surface fluxes predicted by irradiated models. The posterior probability distributions for these models are shown in Figures \ref{cornerh} and \ref{cornerk}.

Our model of the system in the $H$ band predicts a nightside temperature of the brown dwarf of $1540^{+90}_{-70}$ K and the fraction of flux from the white dwarf absorbed by the brown dwarf as 0.50$\pm$0.06. The equivalent model for the $K_s$ band predicts 1000$\pm 500$ K and 0.80$\pm$0.15.  As the $J$ band eclipse was not observed, we were unable to fit a model to these data, and instead fitted a sine curve to the data to measure the reflection effect as was done in \citet{casewell15} for WD0137-349.

\begin{figure}
\begin{center}
\scalebox{0.55}{\includegraphics[]{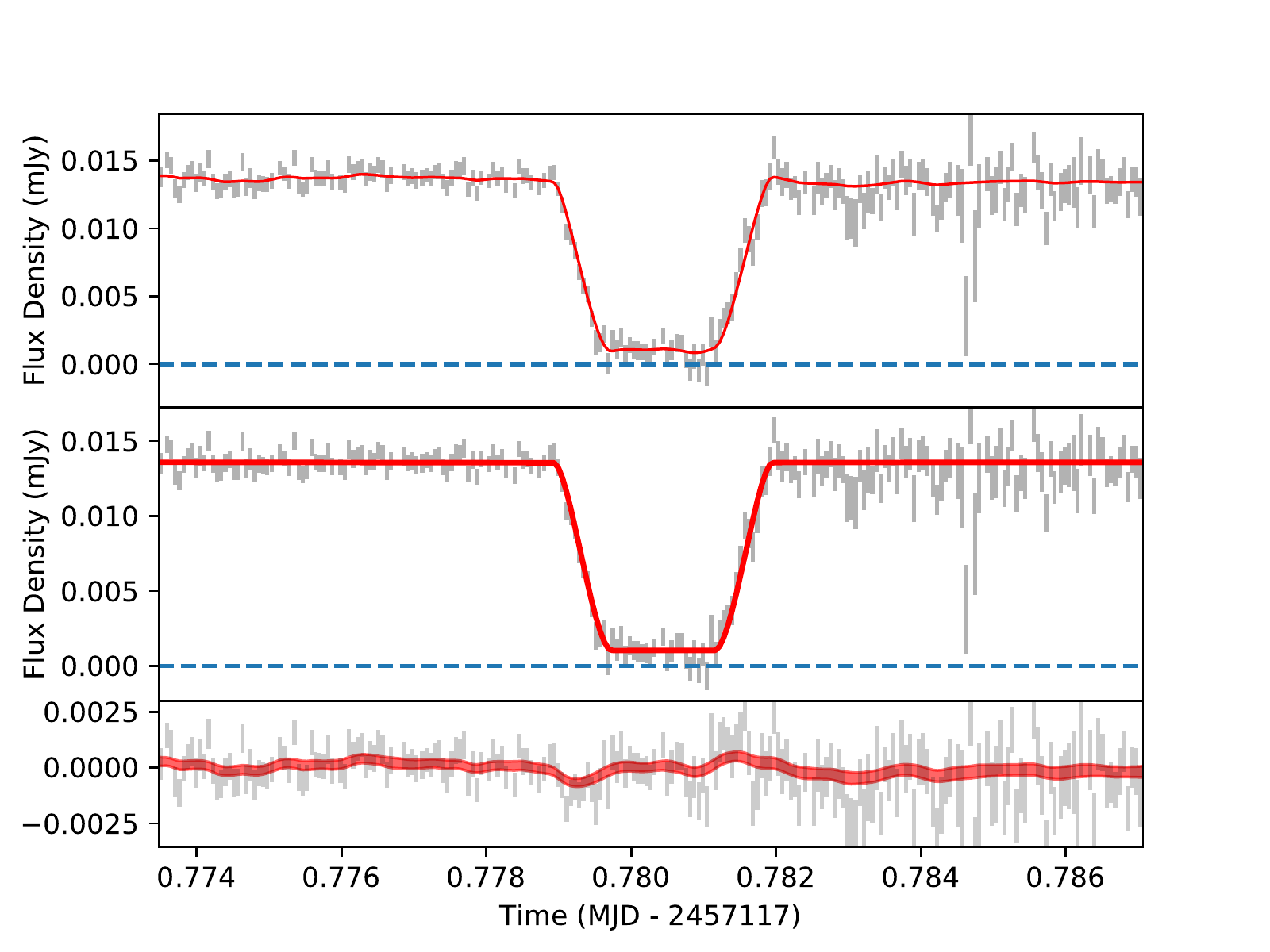}}
\caption{\label{zoom1}$H$ band lightcurve phased on the ephemeris in \citet{beuermann13}, and zoomed in on the eclipse. The top panel shows the raw lightcurve and the binary plus Gaussian process model. The middle panel shows the data with the Gaussian process subtracted, and the binary model alone. The bottom panel shows the residuals to the binary model, and the Gaussian process. The models are plotted with the red solid line, and zero flux is marked by the dotted line.  
}
\end{center}
\end{figure}

\begin{figure}
\begin{center}
\scalebox{0.45}{\includegraphics[]{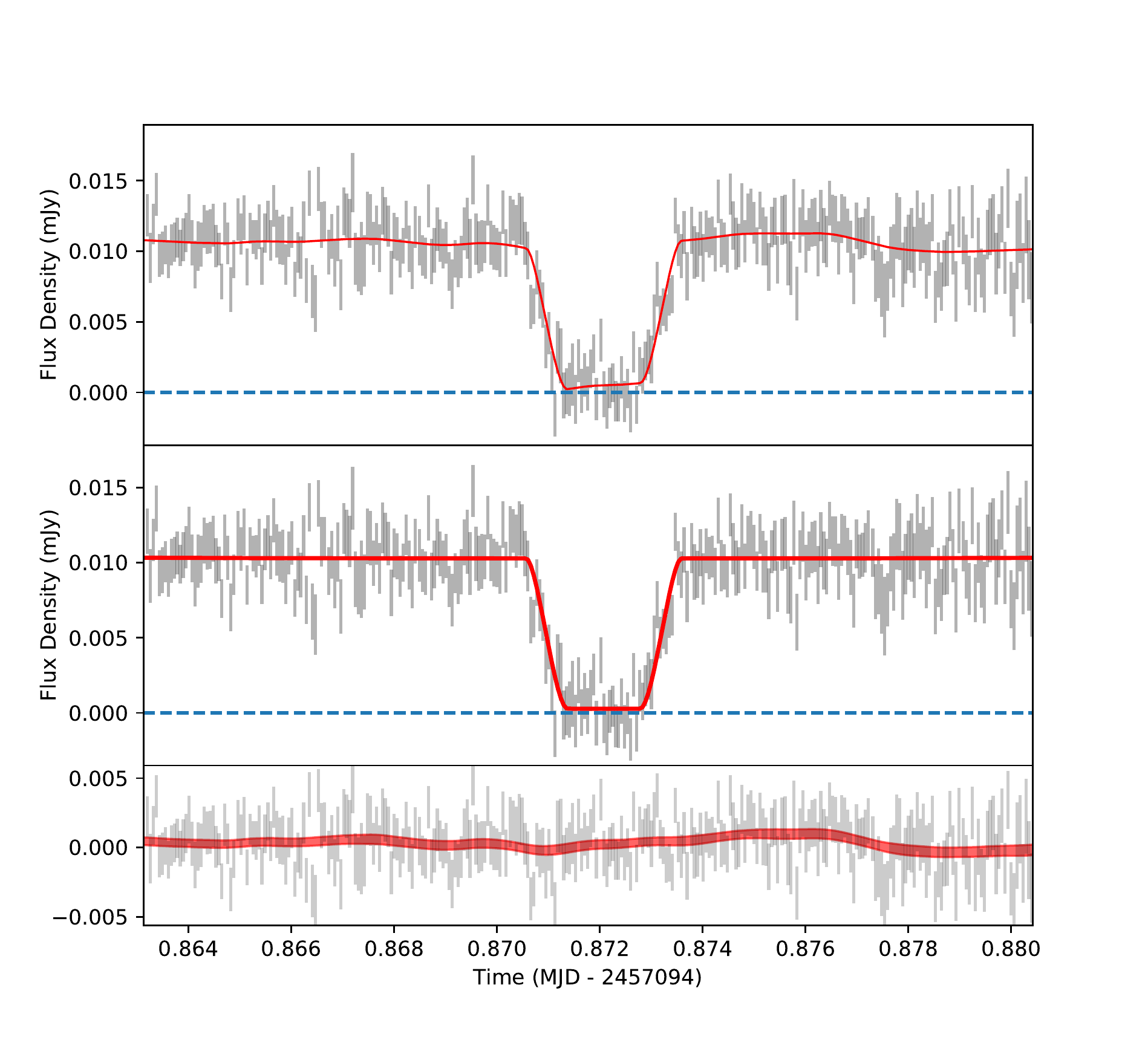}}
\caption{\label{zoom2}$K_s$ band lightcurve phased on the ephemeris in \citet{beuermann13}, and zoomed in on the eclipse. Panels and plot markers are the same as in figure~\protect\ref{zoom1} . 
}
\end{center}
\end{figure}

We detect the primary eclipse of the white dwarf in both the $H$ and $K_s$ data (Figures \ref{zoom1} and \ref{zoom2}). We do not detect the secondary eclipse in any of our data. Our model predicts that the secondary eclipse depth is 0.8 per cent in the $H$ band and 3 per cent in the $K_s$ band, which is smaller than our photometric errors ($\sim$0.05 mags in $H$, and 0.2 mags in $K_s$), and as the secondary eclipse is predicted to last $\sim$ 4 minutes including ingress and egress, we cannot bin our data up to a high enough precision. 

The primary eclipse is total, not unexpected, as brown dwarf radii are typically comparable to that of Jupiter, while white dwarfs have radii similar to that of the Earth, hence all the flux we detect is from the nightside of the brown dwarf at this point. This flux is significantly non-zero in the $H$-band, making this the first direct detection of the dark side of an irradiated brown dwarf. The flux in the $K_s$ band is consistent with zero, which is reflected in our large uncertainties on the $K_s$ brightness temperature. Although our model has calculated an average nightside temperature in the  $K_s$ band, we have chosen to give the nightside an upper limit of 1500 K to reflect the zero flux.

In addition to the detection of the night-side of the brown dwarf, we are also able to calculate the magnitude of the day-side of the brown dwarf due to the reflection effect in the system, causing sinusoidal variations as the tidally locked brown dwarf orbits the white dwarf. The semi-amplitude of this variability is 0.0019$\pm$0.0003 mJy in the $H$ band, and 0.0039$\pm$0.0006 mJy in the $K_s$ band. This variability is slightly larger than that detected for the WD0137-349AB system \citep{casewell15} which has a similar period, but a hotter, and less massive white dwarf (T$_{\rm eff}=$16500 K, M=0.4M$_{\odot}$  \citealt{maxted06}), but the errors are large on these measurements.

We also used the \textsc{molly} software package to search for any emission lines from the brown dwarf in the 28 UVB and VIS XSHOOTER spectra used to measure the radial velocity in \citet{littlefair14}. We did not detect H$\alpha$ emission, as is seen for WD0137-349B \citep{maxted06},  or any other emission lines as were detected by \citet{longstaff17} for the same system.  As SDSS1411+2009 is 3 magnitudes fainter in the optical than the WD0137-349 system, we phase binned the data and combined the spectra in phase, but still did not detect any emission features from the brown dwarf.  The data from the NIR arm of XSHOOTER are of not good enough quality to be used in any analysis.

\section{Discussion}

We calculated brightness temperatures for the dayside of the brown dwarf for the $J$ band using a model white dwarf spectrum and the method detailed in \citet{casewell15}.  For the $H$ and $K_s$ bands where we have models of the system from \textsc{lcurve} we generated a temperature map of the surface of the brown dwarf as was done in \citet{hernandez16}. The average dayside and nightside temperatures are reported in Table \ref{brightness}, although, from the surface map of the brown dwarf we were also able to model the maximum and minimum temperatures present across the surface. These temperatures  had a maximum of 1940$\pm$70 K in the $H$ and 2000$\pm$150 K in the $K_s$ bands, and a minimum of 1530$\pm$90 K and 950$\pm$500 K in the $H$ and $K_s$ bands respectively.  

\begin{figure}
\begin{center}
\scalebox{0.3}{\includegraphics[]{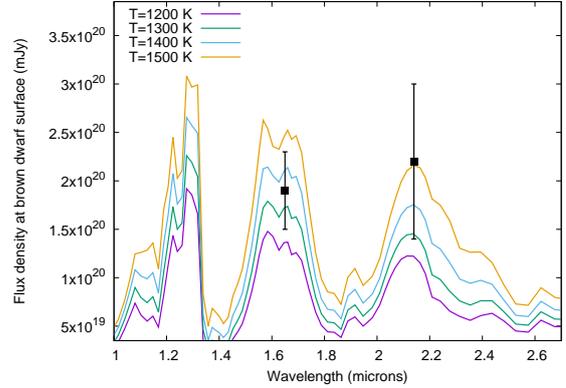}}
\caption{\label{brightnessfig}Dayside average surface flux densities of the brown dwarf (boxes) are shown with irradiated brown dwarf models of effective temperatures: 1200 K, 1300 K, 1400 K and 1500 K.}

\end{center}
\end{figure}

We have generated irradiated brown dwarf models using the atmospheric structure model of \citet{marley99}, \citet{marley02} and \citet{fortney05} using the log g from \citet{littlefair14} and intrinsic effective temperatures (the temperature the brown dwarf would have in the absence of the white dwarf) ranging from  500 K to 1500 K (Figure \ref{brightnessfig}).  The white dwarf irradiation was modelled using a 13 000 K black body at the appropriate separation. We have chosen to use surface flux densities in displaying these data, as this removes any uncertainties associated with the radius of the brown dwarf and the distance to the system. While the dayside $H$ and $K_s$ band fluxes are consistent with an irradiated brown dwarf of 1300 K, it is clear that the dayside $K_s$ flux also encompasses temperatures much hotter than 1500 K (the hottest model plotted). This is consistent with our findings in \citet{casewell15}, where the $K_s$ band was much brighter than the models predicted.

\begin{figure}
\begin{center}
\scalebox{0.3}{\includegraphics[]{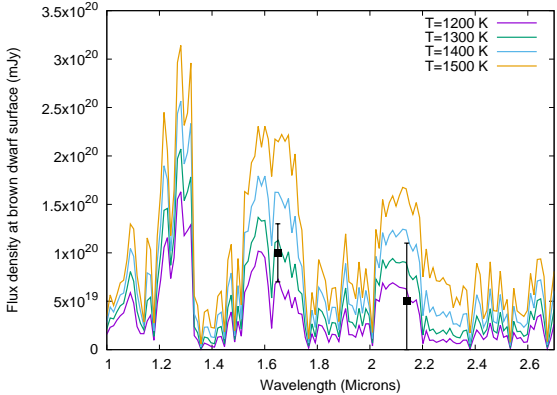}}
\caption{\label{nightside} Nightside average surface flux densities of the brown dwarf (boxes) are shown with non-irradiated, cloud free brown dwarf models of the same effective temperatures as in Figure \ref{brightnessfig}.}
\end{center}
\end{figure}

It can be seen that SDSS1411-2009B has an average difference in day-night side temperatures of 93$\pm$12 K in the $H$ band, and a 360$\pm$80 K day-night difference in the $K_s$ band. As these measurements are derived from the \textsc{lcurve} model, they take into account the errors on the radii and the correlated errors relating to the distance to the binary. The distance from Gaia DR2 is 177$\pm$5 pc \citep{gaia}, compared to 190$\pm$8 in \citealt{littlefair14}. These distances agree to within 1.5 $\sigma$. Surface flux densities for the brown dwarf were derived from the brightness temperatures of each element using the Planck curve.  To compare the nightside fluxes with models, we used non-irradiated cloud free brown dwarf models, again using the atmospheric structure of \citet{marley99}, \citet{marley02} and \citet{fortney05}.  These models, and the nightside fluxes can be seen in Figure \ref{nightside}. Both the $H$ and $K_s$ bands are consistent with T$_{\rm eff}$= 1300 K.  This raises an interesting conundrum, as the estimated T$_{\rm eff}$ of the brown dwarf using the radius from the lightcurves and the mass from the radial velocity solution combined with evolutionary models of \citet{baraffe03} is $\sim$ 800 K \citep{littlefair14}.

In comparison with a similar system, both SDSS1411-2009B and WD0137-349B have similar brightness temperatures (within the errors) in the $H$ band on both the day and night sides, although the nightside of WD0137-349B is an upper limit, and  the dayside temperature is not well constrained at 1585$\pm$329 K compared to 1730$\pm$70 K for SDSSJ1411-2009B.  The nightside of SDSS1411-2009B in the $H$ band appears to be hotter in than that of WD0137-349B, despite SDSSJ1411-2009B being of a later spectral type, but the errors on the upper limit mean we can not state this conclusively.

In the $K_s$ band, the dayside of WD0137-349B (2015 K)is hotter than the dayside of SDSSJ1411-2009B (1620 K), as would be expected for a brown dwarf orbiting a hotter white dwarf in a shorter orbit.  The nightside brightness temperatures of both objects have large errors, however, as with the $H$ nightside measurements, they may be similar temperatures.

\begin{table*}
\caption{\label{brightness}Average brightness temperatures and apparent magnitudes for the system. The errors given are the standard 68 per cent confidence interval.}
\begin{center}
\begin{tabular}{ccccc}
\hline
Waveband &{Magnitude (WD+BD) }& Magnitude (WD)&\multicolumn{2}{c}{Brightness Temperature (K)}\\
& Dayside& & Dayside & Nightside\\ 
\hline
$J$&17.96$\pm$0.04&18.02&1715$^{+95}_{-131}$&-\\
$H$&17.80$\pm$0.04&18.18&1730$\pm$70&1530$^{+90}_{-70}$\\
$K_s$&18.11$\pm$0.10&18.27&1620$\pm$160&1500\\
\hline
\end{tabular}
\end{center}
\end{table*}

Despite the white dwarf in SDSSJ1411+2009
being $\sim$ 3500 K cooler than the white dwarf in
WD0137-349B, there is not a large difference in the SED of the irradiated brown dwarfs in these systems. 
WD0137-349B emits much more strongly in the ultraviolet (by a factor of $\sim$10) than SDSS1411-2009A does, although the peak of the white dwarf SED is approximately at the same wavelength in both cases. This is likely to be the explanation for the lack of emission lines seen in the atmosphere of SDSSJ1411-2009B. The lack of UV irradiation means SDSSJ1411+2009 is unlikely to have a chromosphere, similar to that suggested for WD0137-349B by \citet{longstaff17}.   This is also suggested by the lack of H$\alpha$ emission lines in the optical spectra.  However, despite this lack of emission lines, the same brightening is seen in the $K_s$  for both WD0137-349B and SDSSJ1411-2009B. 


Our nightside brightness temperatures for SDSSJ1411-2009B indicate that in the absence of any heat transport, the T$_{\rm eff}$ of the brown dwarf is 1300 K. Our \textsc{lcurve} modelling of these lightcurves gives an absorb parameter (the fraction of flux from the white dwarf absorbed by the brown dwarf)  of 0.50$\pm$0.06 in the $H$ band  and 0.80$\pm$0.15 in the $K_s$ band. These parameters mean that if only absorption and reprocessing within the brown dwarf atmosphere is important, SDSS1411J-2009B must be absorbing 50 per cent of the $H$ band flux and 80 per cent of the $K_s$ band flux, in order to produce the dayside brightness temperatures. 

However, the brown dwarf effective temperature as estimated from the mass and radius is 800 K \citep{littlefair14}. If this is the true effective temperature of the brown dwarf, were it an isolated object,  then the absorb parameters must be even higher in order to produce enough heat transport to heat the nightside to 1300 K. The absorb parameter for the $K_s$ band is already close to 100 per cent though, which would indicate there is poor energy circulation around the brown dwarf, supported by the 200 K day-nightside difference in the $H$ band.

An additional factor that would affect estimates of temperature and energy circulation, may be fluorescence or emission within the brown dwarf atmosphere. We suggested this is present in WD0137-349B \citep{casewell15}, again causing brightening in the $K_s$ and 4.5 micron bands. If this emission is present, it will increase the dayside flux, particularly in the $K_s$ band, meaning that the absorb parameter is artificially high. In particular it would mean that the brown dwarf needs to absorb a smaller fraction of flux in order to heat the nightside. This scenario is also potentially consistent with a lower T$_{\rm eff}$ of the brown dwarf. Emission from the dayside has artificially increased the flux, leading to an overestimate of the effective temperature.

Observations of Kelt-1b, a T2 dwarf orbiting a main sequence star \citep{siverd12}, seem to support the hypothesis of UV-induced brightening in the $K_s$ band.  Kelt-1b, orbiting a 6500 K F5V star lacks the intense UV irradiation of the white dwarf irradiated systems, and does not show this brightening. Indeed eclipse measurements suggest that this object fits very well with a field dwarf template \citep{croll15,beatty17}.  

The only way we can, however, confirm this hypothesis of UV-induced  emission is by obtaining spectrophotometry of SDSSJ1411-2009B with JWST. This would allow us to determine if at $K_s$ and 4.5 microns the brown dwarf looks like an isolated field object on the dayside, or whether UV-induced emission lines are present.

\section{Conclusions}
We have observed the close, post-common envelope binary SDSS1411+2009 with HAWK-I in the $JHK_s$ bands, and have directly detected the brown dwarf in the $H$ and $K_s$ bands as it eclipses its white dwarf companion.  We have determined the brightness temperatures for the day and night-sides of the brown dwarf and measure a temperature difference of only $\sim$200 K, compared to $\sim$500 K for WD0137-349B, a system with a similar period, but a hotter white dwarf primary.   From comparing the surface fluxes to models of irradiated and non-irradiated brown dwarfs, we also determine that in general, the models indicate the brown dwarf is consistent with  T$_{\rm eff}$=1300 K, but that the mass and radius suggest an effective temperature that is much lower. As the brown dwarf is already absorbing almost all the emission from the white dwarf in the $K_s$ band, this discrepancy suggests that an additional mechanism is making the $K_s$ band brighter. This mechanisms may  similar to that suggested in WD0137-349B, hinting this may be a common trait in these systems, and may be due to  photochemistry.

\section{Acknowledgements}
\noindent We thank Detlev Koester for providing the white dwarf models. This work is based on observations made with ESO Telescopes at the La Silla Paranal Observatory.  This work also makes use of the white dwarf models  from Pierre Bergeron:
$\sim$bergeron/CoolingModels. 
S.L. Casewell acknowledges support from the University of Leicester College of Science and Engineering. SPL is supported by STFC grant ST/M001350/1, and TRM is supported by STFC grant ST/L000733. SGP acknowledges the support of the Leverhulme Trust.
\bibliographystyle{mnras}

\bibliography{wd0137_bib}

\appendix
\section{Posterior probability distributions}

\begin{figure*}
\begin{center}
\scalebox{0.3}{\includegraphics[]{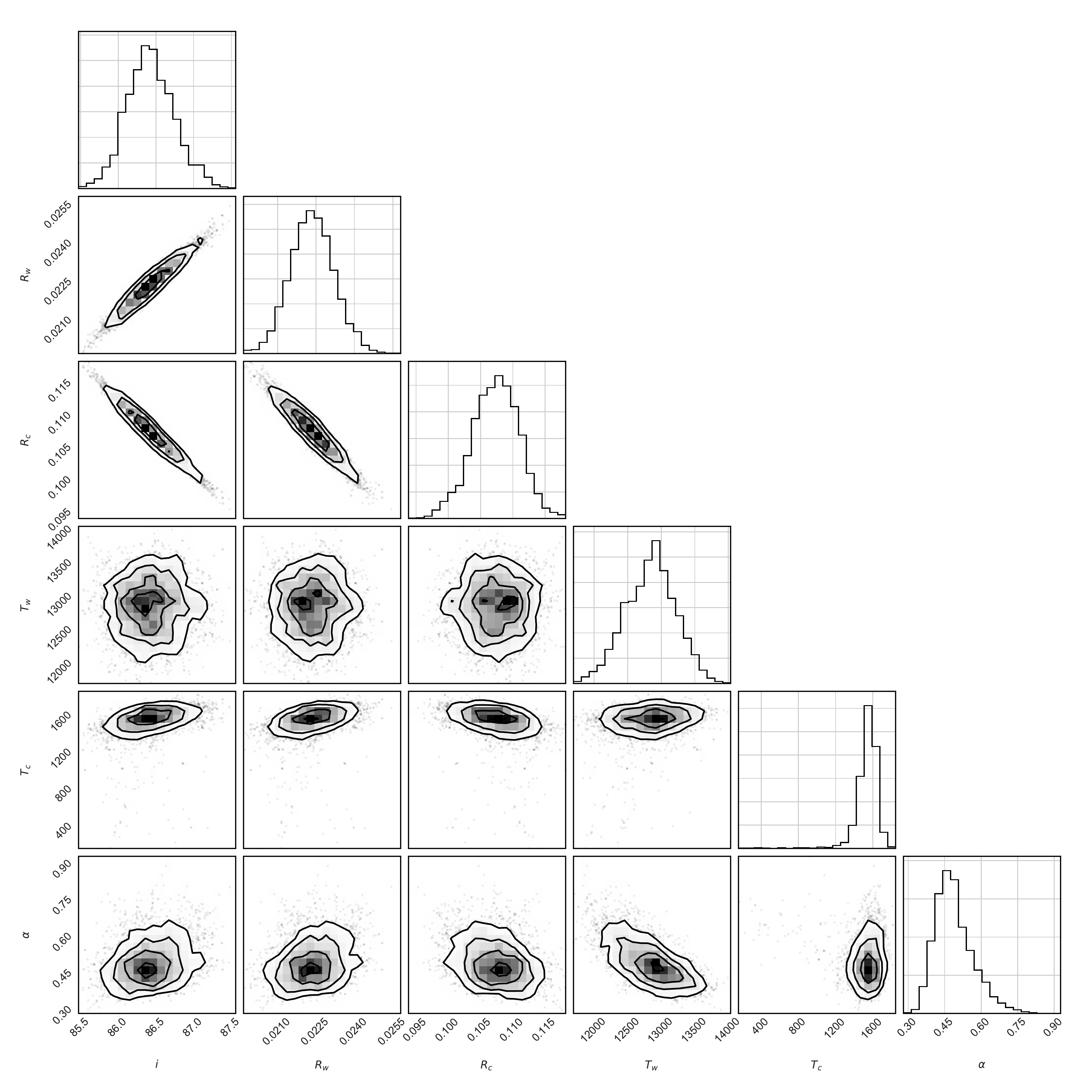}}
\caption{\label{cornerh}Posterior probability distributions for model parameters obtained through fitting the $H$ band lightcurve. See Section 3 for details of the model used. Grey-scales and contours illustrate the joint probability distributions for each pair of parameters, whilst histograms show the marginalised probability distribution for each individual parameter.}
\end{center}
\end{figure*}

\begin{figure*}
\begin{center}
\scalebox{0.3}{\includegraphics[]{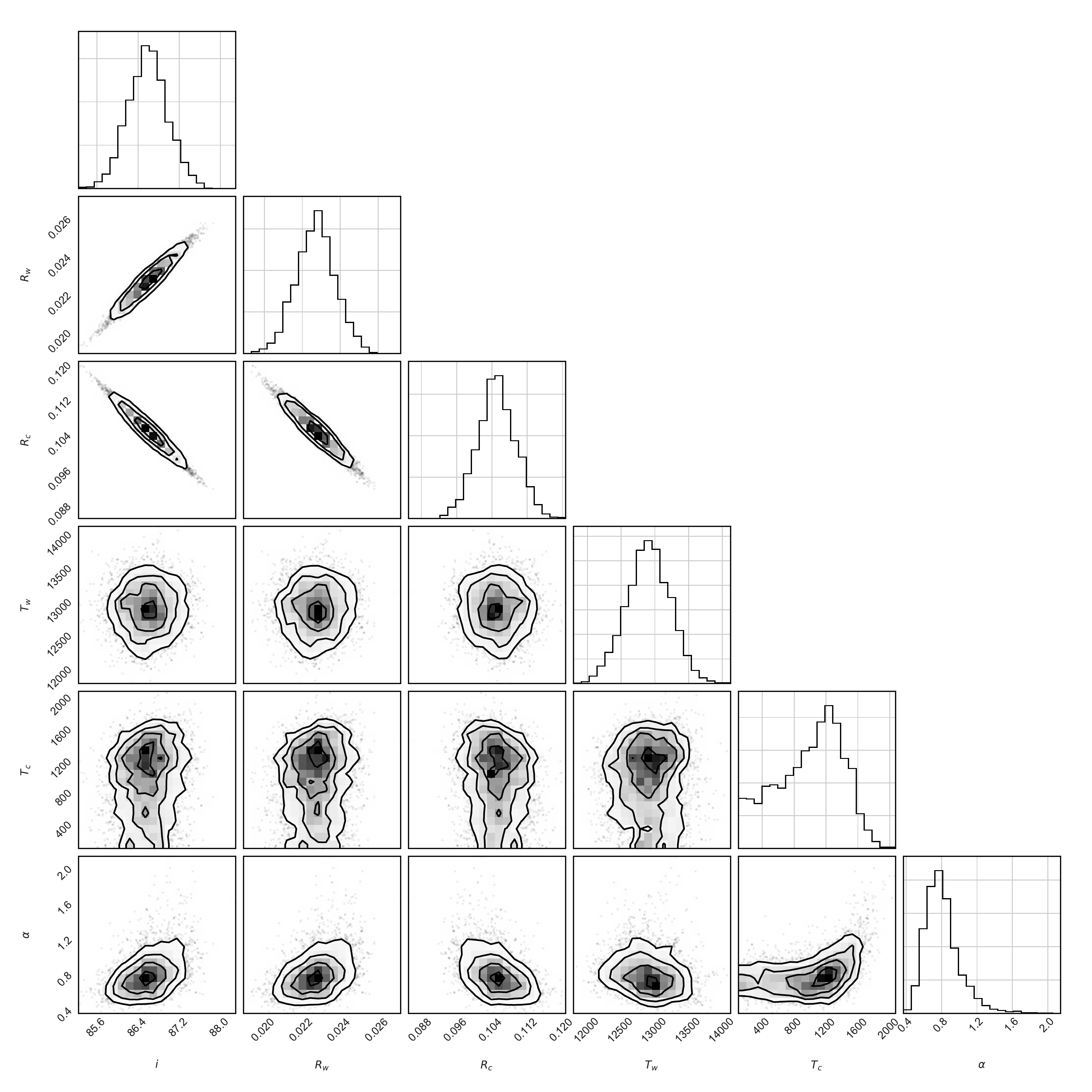}}
\caption{\label{cornerk}Same as for Figure \ref{cornerh} in the $K_s$ band.}
\end{center}
\end{figure*}

\label{lastpage}
\end{document}